\documentclass[twocolumn,prl, superscriptaddress]{revtex4-1}
\usepackage[latin9]{inputenc}
\usepackage{xcolor}
\usepackage{pdfcolmk}
\usepackage{amsmath}
\usepackage{amssymb}
\usepackage{graphicx}
\PassOptionsToPackage{normalem}{ulem}
\usepackage{ulem}
\usepackage[unicode=true,pdfusetitle,
 bookmarks=true,bookmarksnumbered=false,bookmarksopen=false,
 breaklinks=false,pdfborder={0 0 1},backref=false,colorlinks=false]
 {hyperref}
\hypersetup{
 colorlinks,linkcolor=red,citecolor=blue}

\makeatletter

\providecolor{lyxadded}{rgb}{0,0,1}
\providecolor{lyxdeleted}{rgb}{1,0,0}

\makeatother

\begin{document}

\title{Quantum transduction with adaptive control}

\author{Mengzhen Zhang}

\affiliation{Departments of Applied Physics and Physics, Yale University, New
Haven, CT 06520, USA}

\affiliation{Yale Quantum Institute, Yale University, New Haven, CT 06520, USA}

\author{Chang-Ling Zou}

\affiliation{Departments of Applied Physics and Physics, Yale University, New
Haven, CT 06520, USA}

\affiliation{Yale Quantum Institute, Yale University, New Haven, CT 06520, USA}

\affiliation{Key Laboratory of Quantum Information, University of Science and
Technology of China, CAS, Hefei, Anhui 230026, China}

\author{Liang Jiang{*}}

\affiliation{Departments of Applied Physics and Physics, Yale University, New
Haven, CT 06520, USA}

\affiliation{Yale Quantum Institute, Yale University, New Haven, CT 06520, USA}
\begin{abstract}
Quantum transducers play a crucial role in hybrid quantum networks.
A good quantum transducer can faithfully convert quantum signals from
one mode to another with minimum decoherence. Most investigations
of quantum transduction are based on the protocol of direct mode conversion.
However, the direct protocol requires the matching condition, which
in practice is not always feasible. Here we propose an adaptive protocol
for quantum transducers, which can convert quantum signals without
requiring the matching condition. The adaptive protocol only consists
of Gaussian operations, feasible in various physical platforms. Moreover,
we show that the adaptive protocol can be robust against imperfections
associated with finite squeezing, thermal noise, and homodyne detection.
It can be implemented to realize quantum state transfer between microwave
and optical modes.
\end{abstract}


\maketitle
\textit{Quantum transducers} (QT) can convert quantum signals from
one bosonic mode to another, which may have different frequencies,
polarizations, or even mode carriers. QT enables quantum information
transfer between different physical platforms, which is crucial for
hybrid quantum networks \citep{kimble_quantum_2008,duan_colloquium_2010}.
There have been significant advances toward quantum state transfer
between different bosonic systems, such as conversion between microwave
and mechanical/spin-wave modes \citep{Palomaki13,ZhangX14,Tabuchi15},
between optical and mechanical/spin-wave modes \citep{Lukin03,Hammerer10,Safavi-Naeini2011,Aspelmeyer14},
and etc. Motivated by the hybrid quantum networks with optical quantum
communication and microwave quantum information processing, recently
there are experimental demonstrations of coherent conversion between
microwave and optical signals with decent conversion efficiencies
\citep{andrews_bidirectional_2014,vainsencher_bi-directional_2016,Fong14},
but the signal attenuation and added noise still prevent us from achieving
quantum transduction between microwave and optical modes. 

\begin{figure}[t]
\includegraphics[width=0.95\columnwidth]{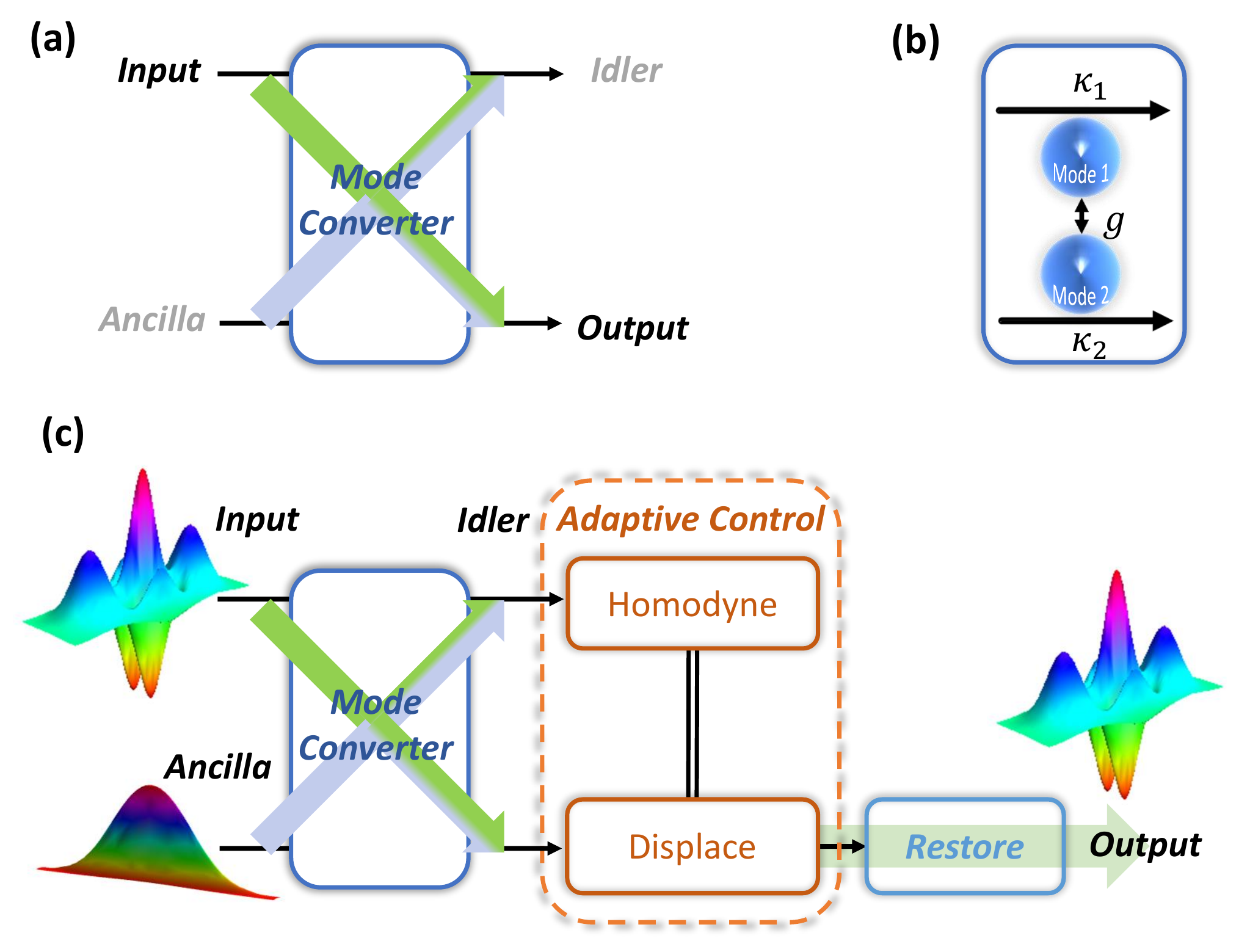}

\caption{Schematic of the direct and adaptive protocols. (a)\textbf{ }The direct
protocol injects quantum signals to the input port and retrieves them
from the output port of the mode converter. (b) A simple mode converter
has bilinear coupling $\hat{H}=\left(ga_{1}^{\dagger}a_{2}+g'a_{1}^{\dagger}a_{2}^{\dagger}+h.c\right)$
between two internal modes $a_{1}$ and $a_{2}$ with coupling strengths
$g$ and $g'$, and external coupling strength $\kappa_{1}$ and $\kappa_{2}$.
(c)\textbf{ }The adaptive protocol injects not only quantum signals
to the input port, but also squeezed vacuum to the ancilla port of
the mode converter. The adaptive control (dashed orange box) performs
a displacement operation to the output port conditioned on the homodyne
detection of the idler port. Up to a unitary recovery operation (cyan
box), quantum signals can be retrieved from the output port. }

\label{Fig1}
\end{figure}

Most investigations of quantum transduction are based on the direct
quantum transduction (DQT) protocol. As illustrated in Fig.$\,$\ref{Fig1}(a),
QDT protocol has a simple structure that injects quantum signals to
the input port and retrieves them from the output port of the mode
converter, which can hybridize different modes with enhanced bilinear
couplings betweeen localized modes (Fig.$\,$\ref{Fig1}(b)). The energy
mismatch between the input and output states can be compensated by
parametric processes and stiff pumps \citep{Pelc12,Abdo13b,andrews_bidirectional_2014,vainsencher_bi-directional_2016,PhysRevLett.117.123902}.
Unlike classical signals, quantum signals are vulnerable to both attenuation
and amplification, which irreversibly add noise and induce decoherence.
Hence, DQT protocol requires the \textit{matching condition} (MC)
\textemdash{} the subblock of the scattering matrix associated with
the input and output ports should be equivalent to the identity matrix
\textemdash{} so that every excitation entering the input port can
be faithfully converted into an excitation exiting the output port,
without affecting other ports \citep{Safavi-Naeini2011,wang_using_2012,kurizki_quantum_2015}.
In practice, however, MC is not always feasible, due to limited tunability
of device parameters \citep{RevModPhys.86.1391} and undesired parametric
conversion processes \citep{andrews_bidirectional_2014}. For small
deviation from MC, we may use quantum error correction to actively
suppress the noise and restore the encoded quantum information \citep{PhysRevA.59.2631,PhysRevA.64.012310,PhysRevLett.111.120501,1367-2630-16-4-045014,michael_new_2016,xiang_intra-city_2016,vermersch_quantum_2016,Ofek16}.
Nevertheless, the quantum error correction has limited capability
of correcting errors (e.g., no more than 50\% loss) \citep{PhysRevLett.78.3217}.
Therefore, it is important to develop a quantum transduction protocol
to bypass MC.

In this Letter, we propose the \textit{adaptive quantum transduction
}(AQT) protocol that does not require MC. Adaptive quantum protocols
have been developed for various applications, including quantum teleportation
\citep{Bouwmeester97,furusawa_unconditional_1998}, quantum phase
estimation \citep{Higgins07}, measurement based quantum computation
\citep{Raussendorf01,menicucci_universal_2006}, and quantum error
correction \citep{NC00}, etc. We incorporate the ingredients of adaptive
control to the general design of quantum transducers to bypass MC
as well as boost the performance. 

\paragraph*{Adaptive quantum transduction.}

As illustrated in Fig.$\,$\ref{Fig1}(c), AQT prepares a squeezed vacuum
for the ancilla port, performs homodyne detection at the idler port,
and applies adaptive control to the output conditioned on the homodyne
outcome. Up to a unitary operation, quantum signals can be converted
from the input to output ports. If MC is satisfied, quantum signals
can be perfectly converted with no need of adaptive control, and thus
AQT is reduced to DQT (Fig.$\,$\ref{Fig1}(a)). If MC is not fulfilled,
the mode converter will distribute the quantum signal (gray arrow)
and squeezed vacuum noise (curved arrow) over \textit{both} output
and idler ports. The quantum signal leaks into the environment via
the idler port, while the noise is added to the output. However, the
squeezed vacuum from ancilla port injects a strong and correlated
noise to the anti-squeezing quadratures of the output and idler ports,
so that we may use homodyne detection and adaptive control to cancel
the added noise as well as prevent the signal leakage to the environment.
On the one hand, the homodyne detection measures the anti-squeezed
noises of the idler port without disclosing the information about
the quantum signal, since the idler port is dominated by the large
fluctuation of the anti-squeezed noise. On the other hand, the adaptive
displacement operation conditioned on the homodyne detection completely
removes the correlated anti-squeezing noise of the output port, leaving
the output signal equivalent to the input signal up to a Gaussian
unitary operation. Since there is no assumption of prior-knowledge
of the input signal, the protocol can faithfully convert arbitrary
quantum signal from one mode to another.

Generally, we consider a mode converter that transforms $m$ input
modes and $n$ ancilla modes into $m$ output modes and $n$ idler
modes. AQT protocol will (1) inject squeezed vacuum $\hat{\rho}_{anc}$
to the ancilla modes, (2) perform homodyne measurement $\hat{\Pi}_{\eta}$
for the idler modes with outcome $\mathbf{\eta}\in\mathbb{R}^{n}$,
and (3) apply adaptive displacement $\mathcal{D}_{\mathbf{F}\mathbf{\mathbf{\eta}}}$
to the output modes with linearly transformed displacement $\mathbf{F}\mathbf{\mathbf{\eta}}\in\mathbb{C}^{m}$.
For arbitrary input state $\hat{\rho}_{in}$, the output state of
AQT is
\begin{equation}
\hat{\rho}_{out}=\int d\mathbf{\eta}\mathcal{D}_{\text{\ensuremath{\mathbf{F}\mathbf{\eta}}}}\left[\mathrm{tr}_{\mathrm{meas}}(\mathcal{U}_{\mathbf{S}}\left[\hat{\rho}_{in}\otimes\hat{\rho}_{anc}\right]\hat{\Pi}_{\eta})\right],\label{eq:rho}
\end{equation}
where $\mathcal{U}_{\mathbf{S}}$ is the Gaussian unitary operation
\citep{weedbrook_gaussian_2012} from the mode converter, which can
be characterized by a symplectic scattering matrix $\mathbf{S}$ transforming
the input and ancilla modes ($\mathbf{x}$) to the output and idler
modes ($\mathbf{y}$)

\begin{eqnarray}
\left(\begin{array}{c}
\mathbf{y}{}_{b}\\
\mathbf{y}{}_{b'}\\
\mathbf{y}{}_{h}\\
\mathbf{y}{}_{h'}
\end{array}\right) & = & \left(\begin{array}{cccc}
\mathbf{S}_{b,a} & \mathbf{S}_{b,a'} & \mathbf{S}_{b,z} & \mathbf{S}_{b,z'}\\
\mathbf{S}_{b',a} & \mathbf{S}_{b',a'} & \mathbf{S}_{b',z} & \mathbf{S}_{b',z'}\\
\mathbf{S}_{h,a} & \mathbf{S}_{h,a'} & \mathbf{S}_{h,z} & \mathbf{S}_{h,z'}\\
\mathbf{S}_{h',a} & \mathbf{S}_{h',a'} & \mathbf{S}_{h',z} & \mathbf{S}_{h',z'}
\end{array}\right)\left(\begin{array}{c}
\mathbf{x}_{a}\\
\mathbf{x}_{a'}\\
\mathbf{\mathbf{x}}_{z}\\
\mathbf{x}_{z'}
\end{array}\right),\label{eq:S1}
\end{eqnarray}
with $\mathbf{x}_{a}(\mathbf{x}_{a'})$ for all the Q(P)-quadratures
of the input modes, $\mathbf{y}{}_{b}(\mathbf{y}{}_{b'})$ for the
Q(P)-quadratures of the output modes, $\mathbf{x}_{z(z')}$ for the
squeezed (anti-squeezed) quadratures of the ancillary modes, and $\mathbf{y}_{h(h')}$
for the measured (unmeasured) quadratures of the idler modes. MC corresponds
to a special case that the subblock $\begin{pmatrix}\mathbf{S}_{b,a} & \mathbf{S}_{b,a'}\\
\mathbf{S}_{b',a} & \mathbf{S}_{b',a'}
\end{pmatrix}$ of the scattering matrix is equivalent to the identity matrix up
to some symplectic transformation \citep{Safavi-Naeini2011,wang_using_2012,SM},
but here we do not require such a condition for AQT. We may choose
the squeezed and measured quadratures ($\mathbf{x}_{z}$ and $\mathbf{y}_{h}$),
so that the anti-squeezed noise in $\mathbf{x}_{z'}$ can be inferred
from the homodyne detection of $\mathbf{y}_{h}$ associated with an
invertible submatrix $\mathbf{S}_{h,z'}$. We choose the linear transformation
\begin{equation}
\mathbf{F}=\mathbf{F}_{\star}=-\begin{pmatrix}\mathbf{S}_{b,z'}\\
\mathbf{S}_{b',z'}
\end{pmatrix}\left(\mathbf{S}_{h,z'}\right)^{-1},\label{eq:F}
\end{equation}
which can completely remove the anti-squeezed noise from the output
modes. Moreover, for this particular choice of $\mathbf{F}_{\star}$,
the effective scattering matrix between the input and output is
\begin{eqnarray}
\tilde{\mathbf{S}} & = & \begin{pmatrix}\mathbf{S}_{b,a} & \mathbf{S}_{b,a'}\\
\mathbf{S}_{b',a} & \mathbf{S}_{b',a'}
\end{pmatrix}+\mathbf{F}_{\star}\begin{pmatrix}\mathbf{S}_{h,a} & \mathbf{S}_{h,a'}\end{pmatrix},\label{eq:S2}
\end{eqnarray}
which is a symplectic matrix, as shown in Theorem 1 of \citep{SM}.
Unlike general scattering matrices, the symplectic $\mathbf{\tilde{S}}$
implies that the output state (after the adaptive displacement) is
a simple Gaussian unitary transformation of the input state
\begin{equation}
\hat{\rho}_{out}=\mathcal{U}_{\tilde{\mathbf{S}}}\left[\hat{\rho}_{in}\right],
\end{equation}
where $\mathcal{U}_{\tilde{\mathbf{S}}}$ is the Gaussian unitary
operation associated with symplectic $\mathbf{\tilde{S}}$. We can
perfectly restore the original input state by applying a unitary recovery
operation $\mathcal{U}_{\tilde{\mathbf{S}}}^{-1}$ over the output
modes, $\hat{\rho}_{out}\rightarrow\mathcal{U}_{\tilde{\mathbf{S}}}^{-1}\left[\hat{\rho}_{out}\right]=\hat{\rho}_{in}$.
Since AQT protocol works for generic scattering matrix $\mathbf{S}$,
it can bypass MC to achieve perfect conversion of arbitrary quantum
signals.

\paragraph*{Finte squeezing and imperfect homodyne.}

So far, we have assumed the ideal situation with infinite squeezing
and perfect homodyne detection for AQT protocol. In practice, however,
we only have finite squeezing and imperfect homodyne detection. The
finite squeezing can be characterized by $\nu=e^{-2\xi}(2n_{z}+1)$,
depending on the squeezing parameter $\xi$ and thermal noise $n_{z}$
prior to squeezing. In terms of logorithmic unit of decibel, $x\rightarrow10\log_{10}x~(\mathrm{dB})$,
squeezing of $\nu\approx-15\mathrm{dB},-10\mathrm{dB}$ for optical
and microwave modes have been achieved \citep{castellanos-beltran_amplification_2008,PhysRevLett.104.251102},
respectively. The imperfect homodyne detection can be characterized
by $\mu=\frac{1-\eta}{\eta}$,  depending on the detector efficiency
$\eta\le1$ . In terms of decibel \footnote{We can justify the use of decibel for $\mu$. Given an ideal EPR pair,
the imperfect homodyne detection of one mode prepares the other mode
in a squeezed state with squeezing parameter $\mu$.}, we can achieve homodyne detection with achievable imperfection of
$\mu\approx-14\mathrm{dB},-0.1\mathrm{dB}$ for optical and microwave
modes have been demonstrated \citep{fuwa_experimental_2015,PhysRevLett.106.220502,kindel_generation_2016},
respectively. Since these imperfections can be characterized by Gaussian
operations, AQT protocol with imperfections is still a Gaussian channel,
which preserves the Gaussian character of a Gaussian state \citep{weedbrook_gaussian_2012}.
With the choice of $\mathbf{F}=\eta^{-1/2}\mathbf{F}_{\star}$, AQT
protocol combined with the recovery operation $\mathcal{U}_{\tilde{\mathbf{S}}}^{-1}$
is effectively a classical-noise channel \citep{Holevo2007,weedbrook_gaussian_2012},
which transforms the quadratures as $\left(\mathbf{x}_{a},\mathbf{x}_{a'}\right)\rightarrow\left(\mathbf{x}_{a}+\xi,\mathbf{x}_{a'}+\xi'\right)$.
The added noise $\left(\xi,\xi'\right)$ is characterized by a $2m\times2m$
covariance matrix \citep{SM}
\begin{equation}
\mathbf{V}=\nu\mathbf{B}_{\star}\mathbf{B}_{\star}^{T}+\mu\tilde{\mathbf{S}}^{-1}\mathbf{F}_{\star}\mathbf{F}_{\star}^{T}\left(\tilde{\mathbf{S}}^{-1}\right)^{T},
\end{equation}
with $\mathbf{B}_{\star}=\begin{pmatrix}\left(\mathbf{S}^{-1}\right)_{a,h'}\\
\left(\mathbf{S}^{-1}\right)_{a',h'}
\end{pmatrix}\left[\left(\mathbf{S}^{-1}\right)_{z,h'}\right]^{-1}$ . Note that $\mathbf{V}$ vanishes when $\nu\rightarrow0$ (infinite
squeezing) and $\mu\rightarrow0$ (perfect homodyne detection), in
correspondence with for the perfect conversion with the ideal AQT.

\begin{figure}[!t]
\includegraphics[width=0.95\columnwidth]{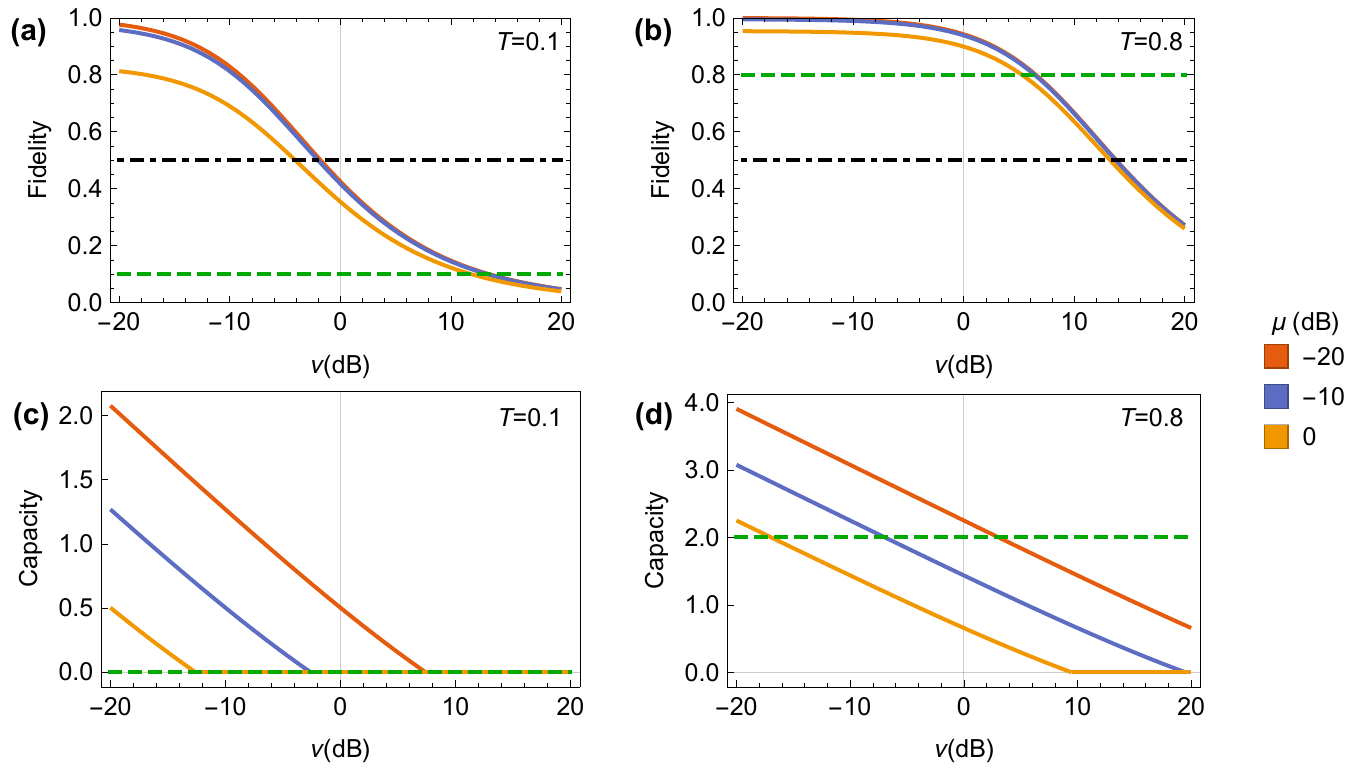}

\caption{Performance of adaptive protocol with imperfect squeezing and homodyne
detection.\textbf{ }(a)\textbf{ }\&\textbf{ }(b),\textbf{ }The average
fidelity of AQT as a function of imperfect squeezing $\nu$, given
imperfect homodyne detection of $\mu=-20,-10$ and $0$dB for transmittance
$T=0.8$ and $T=0.1$, respectively. The dark dotted dashed lines
correspond to the threshold fidelity of $1/2$. The green dashed lines
correspond to the fidelity achieved by DQT. (c)\textbf{ }\&\textbf{
}(d), The quantum channel capacity of AQT for transmittance $T=0.8$
and $T=0.1$, respectively. The green dashed lines correspond to the
channel capacity achieved by DQT. For $T=0.1$, the channel capacity
vanishes for DQT, while AQT can achieve a finite quantum channel capacity
with experimentally feasible $\mu$ and $\nu$.}

\label{Fig2}
\end{figure}

\paragraph*{Performance of adaptive protocol.}

We use two criteria to evaluate the performance of AQT in the presence
of imperfections \textemdash{} (1) the\textit{ average fidelity} between
input and output over uniformly distributed coherent states \citep{braunstein_quantum_2001,fn2}
and (2) \textit{quantum channel capacity} characterizing the amount
of quantum information transmitted \citep{holevo_evaluating_2001,harrington_achievable_2001,pirandola_direct_2009}.
It is sufficient (not necessary) to demonstrate quantum transduction,
if we have above-threshold average fidelity (\textgreater{}1/2) or
quantum channel capacity (\textgreater{}0).

For example, we consider the minimum AQT with $m=1$ input (output)
and $n=1$ ancilla (idler) modes, which is based on a converter with
beam-splitter type coupling {[}e.g., $\hat{H}=g\left(a_{1}^{\dagger}a_{2}+h.c.\right)${]}.
We may simply use the transmittance $T\in\left[0,1\right]$ to characterize
such a converter. Given fixed measurement imperfection ($\mu=0,-10$
or $-20$dB), the average fidelity decreases for larger squeezing
imperfection ($\nu$) as shown in Fig.$\,$\ref{Fig2}(a),(b) for different
$T=0.8$ and $T=0.1$, respectively. For feasible squeezing ($\nu\lesssim0$dB),
AQT can outperform DQT (green curves) and exceed the threshold fidelity
of 0.5 (dark dotted dashed lines) \citep{braunstein_quantum_2001,fn3}.
We can also compute the quantum channel capacity versus squeezing
imperfection as shown in Fig.$\,$\ref{Fig2}(c),(d) for $T=0.8$ and
$T=0.1$, respectively. \footnote{For the mimimum AQT, the quantum channel capacity only depends on
the product of $\mu$ and $\nu$, because the imperfections $\mu$
and $\nu$ adds uncorrelated noise to the two orthogonal quadratures
of the output mode. As detailed in the \citep{SM}, the classical-noise
channel has convariance matrix $\mathbf{V}=\left(1\mp T\right)\text{diag}\left(\frac{1}{T}\nu,\mu\right)$,
for beam-splitter type and two-mode squeezer type converters, respectively.
Since the quantum channel capacity is invariant under squeezing operations,
the channel with $\tilde{\mathbf{V}}=\left(1\mp T\right)^{2}\sqrt{\frac{\mu\nu}{T}}\text{diag}\left(1,1\right)$
has the same quantum channel capacity, which only depends on $\mu\nu$. } When the transmittance is low ($T<0.5$), DQT is an anti-degradable
channel with zero quantum channel capacity \citep{caruso_phys._nodate},
while AQT can still achieve a finite quantum channel capacity when
$\mu\nu<\frac{4}{9\left(T+1/T-2\right)}$ \citep{SM}.

\begin{figure}[!t]
\includegraphics[width=0.95\columnwidth]{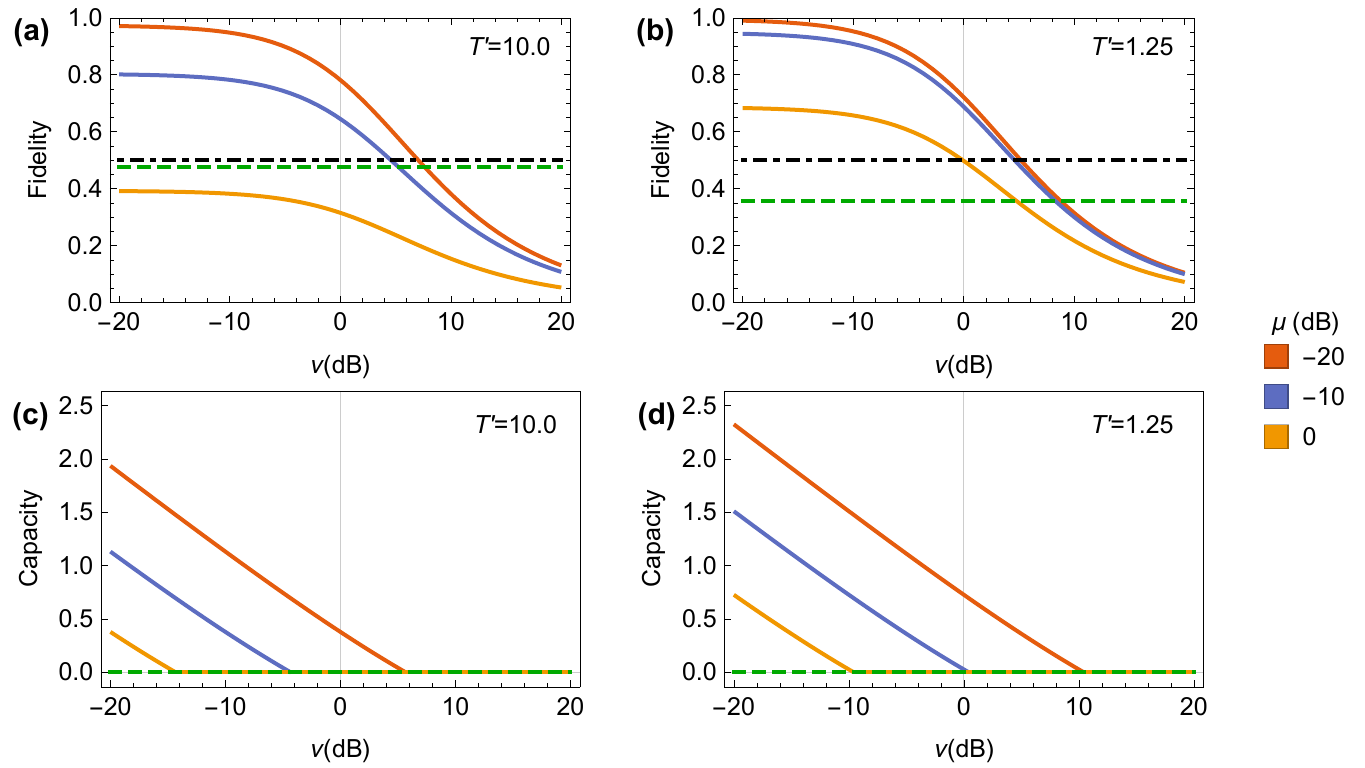}

\caption{Performance of adaptive protocol with imperfect squeezing and homodyne
detection. (a) \& (b),\textbf{ }Input-output fidelity of AQT protocol
averaged over all coherent states as a function of imperfect squeezing
$\nu$, given imperfect homodyne detection of $\mu=-20,-10$ and $0$dB
for transmittance $T'=1.25$ and $T'=10.0$, respectively. The dark
dotted dashed lines correspond to the threshold fidelity of $1/2$.
The green dashed lines correspond to the fidelity achieved by DQT
protocol. (c) \& (d), The quantum channel capacity for transmittance
$T'=1.25$ and $T'=10.0$, respectively. The channel capacity vanishes
for DQT protocol shown by green dashed lines.\label{fig:3}}
\end{figure}

We also investigated AQT based on a converter with two-mode-squeezer
type coupling {[}e.g., $\hat{H}=g\left(a_{1}^{\dagger}a_{2}^{\dagger}+h.c.\right)${]}
characterized by transmittance $T'\in[0,\infty)$. As shown in Fig.$\,$\ref{fig:3}(a),(b),
AQT (dark dotted dashed lines) can have fidelity much higher than
the threshold value of 0.5 with feasible squeezing and homodyne detection,
while DQT (green curves) never exceeds the threshold. Moreover, DQT
with two-mode-squeezer type coupling is always an anti-degradable
channel \citep{caruso_phys._nodate} with zero quantum channel capacity.
Nevertheless, as shown in Fig.$\,$\ref{fig:3}(c),(d), AQT can maintain
a finite quantum channel capacity when $\mu\nu<\frac{4}{9\left(T'+1/T'+2\right)}$
\citep{SM}.

\paragraph*{Discussions.}

AQT can be applied to input with multiple spectral/temporal modes.
For mode converter with a finite bandwidth ($B$), the scattering
matrix will have a deviation depending on $\delta\omega/B$ for modes
with a small detuning $\delta\omega$ from the optimal frequency.
DQT requires $\delta\omega/B\ll1$ to avoid decoherence of quantum
signals even when MC is satisfied. In contrast, AQT can maximize the
capacity of every mode we want to use even when $\delta\omega/B\sim1$,
by using mode-dependent adaptive control.

We have assumed that we have access to all relevant ancilla/idler
ports in our analysis. In practice, we might not have access to all
these ports (e.g. there exist inaccessible intrinsic loss channels)
for mode conversion of quantum signals. Nevertheless, AQT can still
use the accessible ports to maximally restore quantum signals. The
influence of inaccessible ports can be further reduced by optimizing
the conversion matrix $F$, which may inspire us to find more robust
adaptive protocols.

AQT is fundamentally related to other adaptive quantum protocols,
such as continuous variable quantum teleportation. The standard teleportation
scheme needs two ancilla modes in Einstein\textendash Podolsky\textendash Rosen
paradox (EPR) state, two idler modes for homodyne detection, and adaptive
displacement of the output \citep{furusawa_unconditional_1998}. Since
the EPR state can also be obtained by interfering two squeezed ancilla
modes with a balanced beam splitter, the teleportation scheme can
be regarded as a special realization of AQT with $m=1$ input (output)
and $n=2$ ancilla (idler) modes. In addition, AQT can be extended
to the situation of quantum state transfer between $d$-level systems,
by replacing the symplectic mode converter for continuous variable
systems \citep{menicucci_universal_2006} with the Clifford gate coupling
the $d$-level systems. For example, the minimum AQT for $d=2$ corresponds
to the one-bit teleportation circuit \citep{Zhou00}. Moreover, we
may generalize AQT with continuous variable encoding for the input
and ancilla modes, which will enable us to achieve mode conversion
as well as quantum error correction \citep{Knill05nature}. 

In conclusion, we have demonstrated how adaptive control can be a
powerful tool for quantum transduction. In particular, the adaptive
protocol can bypass the matching condition that is vital for previous
direct protocols. The adaptive protocol can boost the averaged fidelity
and quantum channel capacity, while being robust against practical
imperfections. The adaptive approach opens a new pathway of converting
quantum signals among optical, microwave, mechanical, and various
other physical platforms, leading towards the hybrid quantum networks.
\begin{acknowledgments}
We would like to thank Michel Devoret, Konrad Lehnert, Wolfgang Pfaff,
Rob Scheolkopf, Hong Tang for discussions. We also acknowledge support
from the ARL-CDQI, ARO (W911NF-14-1-0011, W911NF-16-1-0563), AFOSR
MURI (FA9550-14-1-0052, FA9550-15-1-0015), ARO MURI (W911NF-16-1-0349),
NSF (EFMA-1640959), Alfred P. Sloan Foundation (BR2013-0049), and
Packard Foundation (2013-39273).
\end{acknowledgments}

\bibliographystyle{apsrev4-1}
\bibliography{references}

\begin{thebibliography}{52}
\bibitem [{fn2()}]{fn2}%
  \BibitemOpen
  \href@noop {} {}\bibinfo {note} {When comparing the input-output fidelity, we
  add an amplification attenuation step to the direct scheme to adjust the
  amplitude of the output mode, to avoid vanishing fidelity.}\BibitemShut
  {Stop}%
\bibitem [{fn3()}]{fn3}%
  \BibitemOpen
  \href@noop {} {}\bibinfo {note} {The experimental requirement of \(
  \tilde{\mathbf{S}} \) operation, especially the amount of squeezing needed to
  restore the quantm signal is \( \nu_{\tilde{\mathbf{S}}} = -10 \log_{10} T
  \)(dB). For \( T=0.1 \), \( \nu_{\tilde{\mathbf{S}}} \) is about 10dB,
  comparable with the required $\nu$ for high fidelity transfer.}\BibitemShut
  {Stop}%
\end{thebibliography}

\end{document}